\documentclass[preprint,aps]{revtex4}
\usepackage{color}
\usepackage{mathtools}
\usepackage{array}
\usepackage{tabularx}
\usepackage{diagbox}
\usepackage{tikz}
\usetikzlibrary{decorations.pathmorphing}
\usepackage{bm}
\usepackage{amsmath,bm,amssymb,amsfonts,dcolumn,color,graphicx,latexsym,epsfig}
\usepackage{bbold}
\usepackage{rsfso}
\usepackage{microtype}
\usepackage{setspace} 
\usepackage{array, booktabs}
\usepackage[utf8]{inputenc}
\usepackage{hyperref}
\hypersetup{
colorlinks=true,
linkcolor=blue,
filecolor=magenta,      
urlcolor=magenta,
citecolor=magenta,
}
\usepackage{multirow}
\usepackage{natbib}
\usepackage{float}
\usepackage{booktabs}
\usepackage{pifont}
\usepackage{mathrsfs}
\usepackage{caption}
\usepackage{caption, threeparttable}
\begin{document}

\title{Spinning test particles in a weak gravitational wave and the memory effect}
\author{Ritwik Acharyya}
\email[Email address: ]{ritwikacharyya@kgpian.iitkgp.ac.in}
\affiliation{Department of Physics, Indian Institute of Technology, Kharagpur 721 302, India}
\author{Shailesh Kumar}
\email[Email address: ]{shailesh@phy.iitkgp.ac.in}
\affiliation{Department of Physics, Indian Institute of Technology, Kharagpur 721 302, India}
\author{Sayan Kar}
\email[Email address: ]{sayan@phy.iitkgp.ac.in}
\affiliation{Department of Physics, Indian Institute of Technology, Kharagpur 721 302, India}

\begin{abstract}
\noindent In this work, we investigate the worldline deviation of spinning test particles in a weak gravitational wave (GW) spacetime. Within the pole–dipole approximation and the Tulczyjew spin supplementary condition, we 
consider the modified deviation equation, which contains the standard geodesic deviation term and an additional contribution generated by the Mathisson-Papapetrou-Dixon (MPD) force. We demonstrate how the deviation equation can be solved under viable approximations to yield a new expression for the change in the deviation vector in terms of the perturbation $h_{ij}$, its time derivative, the spin vector components and the longitudinal and transverse components of the deviation. The spin-curvature coupling is shown to play a crucial role in controlling the change in deviation. We then use this result to investigate the  changes in the form and expressions of the net GW signal as well as our understanding of GW memory, in the presence of spin. We conclude with
our attempts on the measurability of this spin-induced effect 
\emph{vis-a-vis} present GW observations.

\end{abstract}

\pacs{}

\maketitle


\section{Introduction}
\noindent Observations of gravitational waves (GWs) offer new avenues for understanding strong-field effects and the dynamical regime of gravity, thereby enabling precision tests of general relativity (GR) and beyond, together with their astrophysical implications. Following the first direct detection of GW \cite{LIGOScientific:2016aoc}, numerous observations by the LIGO–Virgo–KAGRA network \cite{2025arXiv250907348T,kw5g-d732,LIGOScientific:2017bnn,LIGOScientific:2016sjg} have led to detailed studies on compact binary mergers, rigorous tests of GR in the strong-field regime \cite{PhysRevX.13.011048,LIGOScientific:2016vlm,LIGOScientific:2018dkp,2025arXiv251001001L}, insights into neutron-star physics \cite{2012PhRvL.108a1101B}, and new aspects in cosmology \cite{Palmese_2023,2026arXiv260422731S,2025SCPMA..6910401B}. The interaction of GWs with matter is conventionally described using freely falling test particles which follow geodesic motion. However, realistic probes may carry intrinsic angular momentum associated with the test body. As a result, their dynamics would be governed by the Mathisson–Papapetrou–Dixon (MPD) equations \cite{Mathisson:1937zz,Corinaldesi:1951pb,10.1098/rspa.1951.0200,Dixon:1964cjb,Dixon:1970zz,10.1098/rspa.1970.0020, Tulczyjew}, which incorporates a coupling between spin and spacetime curvature. The spin–curvature interaction, appearing in the MPD equations, gives rise to additional degrees of freedom beyond geodesic motion, which have been investigated in the context of diverse astrophysical systems \cite{PhysRevD.54.3762, PhysRevD.61.024005, 10.1046/j.1365-8711.1999.02754.x, PhysRevD.102.024041, PhysRevD.105.084031, Skoupy:2021asz}. Along this line, previous studies by Mohseni and collaborators showed that the spin–curvature coupling in GW spacetimes leads to nontrivial spin contributions and quantifiable deviations from the usual geodesic trajectories \cite{Mohseni:2000re}, implying that spinning particles could act as sensitive probes of gravitational radiation. Recently, worldline deviation equations for spinning particles in a weak-field have been formulated within the MPD framework \cite{Bini_2017, 2004PhLB..587..133M}, providing a systematic approach to characterize the relative motion of nearby spinning bodies. Since GWs reaching ground-based detectors are sufficiently weak, their interaction with matter is accurately captured by linearized gravity, which makes weak plane-wave spacetimes an ideal setting for examining spin-dependent effects analytically. 

\noindent On the other hand, beyond the familiar oscillatory stretching and squeezing of freely falling test masses, GWs can also produce permanent, non-oscillatory changes in the relative configuration of detectors. This phenomenon, known as the GW memory effect, was first identified by Zel'dovich and Polnarev \cite{1974SvA....18...17Z} as a permanent displacement between initially comoving test particles induced by a burst of gravitational radiation. Braginskii and Grishchuk later termed it as ``memory" \cite{1985ZhETF..89..744B}. Subsequently, Christodoulou discovered a nonlinear contribution to the memory arising from the energy carried by GWs themselves \cite{PhysRevLett.67.1486}, while Thorne provided its physical interpretation in terms of the energy flux radiated to null infinity \cite{PhysRevD.45.520, PhysRevD.44.R2945}. For a detailed review on various aspects of GW memory, see \cite{2010CQGra..27h4036F, PhysRevD.80.024002}. 

\noindent In recent years, GW memory has been studied extensively in a variety of theoretical, astrophysical, and observational settings, including 
proposals for persistent GW-induced observables and their implications for GW detection \cite{2024CQGra..41m5012B, Bhattacharjee_2021, 2026EPJP..141..267A, 2025PhRvD.112d4066A, Singh:2025hsp, 2019PhRvD..99h4044F, 2020PhRvD.101j4033F, Cunningham:2024dog, Zhang_2017, PhysRevD.99.024031, 2022EPJP..137..418C, Inchauspe:2024ibs}. It has also been shown that the memory effect is closely connected to the asymptotic structure of gravity. Its relation to asymptotic symmetries and soft-graviton theorems forms an important aspect of the infrared structure of GR \cite{Strominger:2013jfa, strominger2014gravitationalmemorybmssupertranslations, strominger2018lecturesinfraredstructuregravity, Blanchet_2023, Strominger:2014pwa, PhysRevLett.116.231301, Strominger:2017zoo, PhysRevD.99.084044, PhysRevD.101.104033, Grant_2024, Bhattacharjee:2019jaf, Bhattacharjee:2020lgt, DeLuca:2024cjl}. Recent work has further explored these connections in numerical relativity and in local descriptions of memory, while emphasizing their relevance for GW observations and tests of GR \cite{2022PhRvD.106h4029M, Mitman_2024, Talbot:2018sgr, Hubner:2019sly, PhysRevD.89.084039, PhysRevD.101.104033, Cogez:2026frh, Zosso:2026czc, Bhat:2024cyq, Chakraborty:2019yxn, Chakraborty:2022qvv, Zhang:2017rno, McNeill:2017uvq, Mitman:2020bjf, Khera:2020mcz, Cheung:2024zow, Lasky:2016knh, PhysRevLett.132.241401, Elhashash:2025hqi}. 

\begin{figure}[h]
\centering
\includegraphics[width=0.7\textwidth]{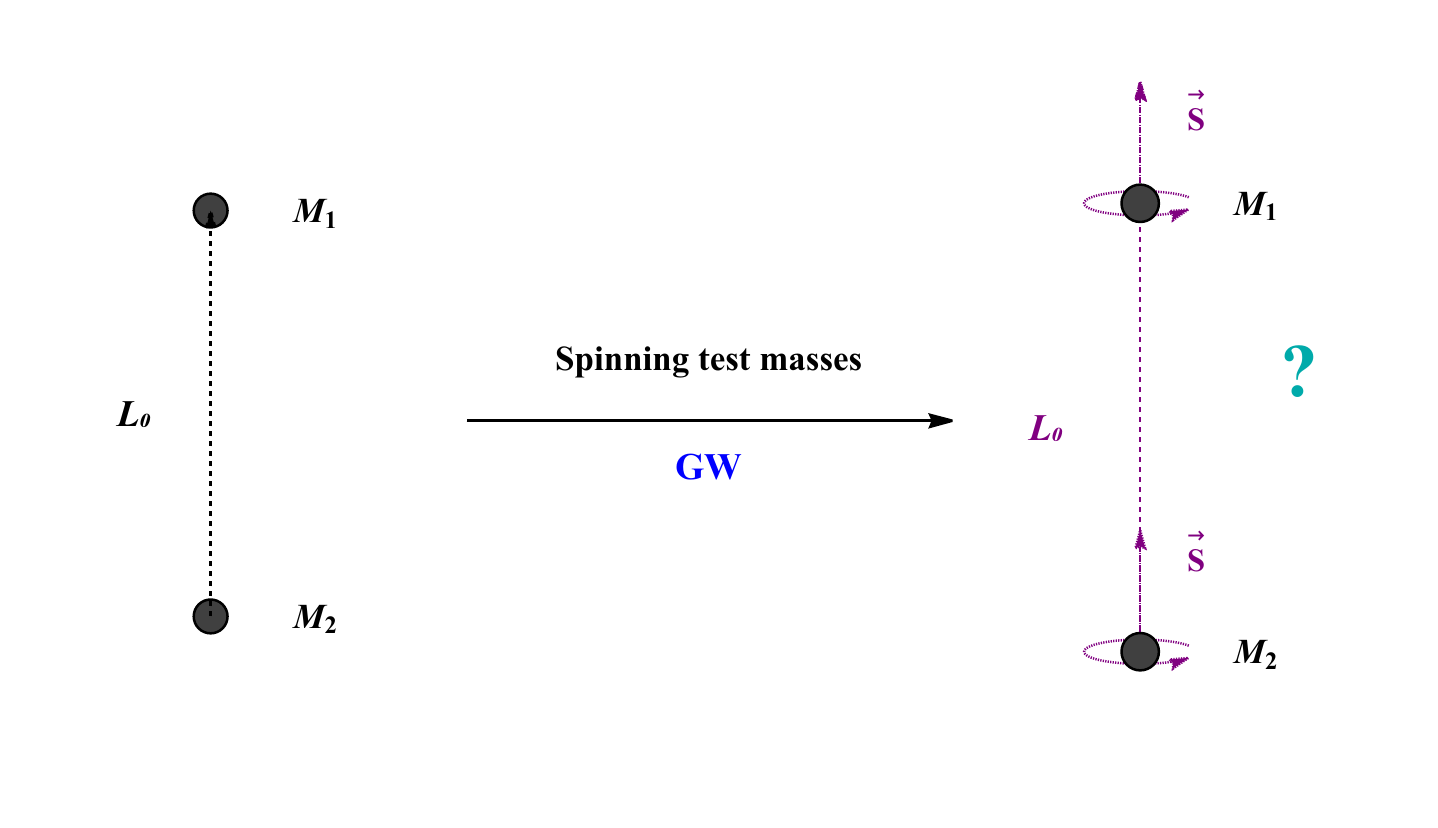}
\caption{Schematic illustration of the response of spinning test masses to a GW. The left panel shows two freely falling test masses, $M_1$ and $M_2$, separated by $L_0$, while the right panel illustrates their generalization to spinning particles. We investigate the modification of the GW response due to spin--curvature coupling and its possible contribution to the memory effect.}
\label{schematic_illustration}
\end{figure}

\noindent However, most studies on memory till date,
involve the response of freely falling test masses which follow geodesics 
and whose relative motion (deviation) is determined solely by the spacetime geometry. A natural generalization of these results would be to 
consider spinning test particles, whose dynamics, as mentioned before, are governed by the MPD equations and involve an additional coupling between the intrinsic spin of the particle and spacetime curvature \cite{Mathisson:1937zz, Corinaldesi:1951pb, 10.1098/rspa.1951.0200, Dixon:1964cjb, Dixon:1970zz, 10.1098/rspa.1970.0020}. This raises the question (schematically shown in Fig. \ref{schematic_illustration}) whether intrinsic spin
while modifying the GW response could leave a measurable imprint on memory.

\noindent Motivated by the above, we study the response of spinning test particles to a weak plane GW, focusing on the modification induced by spin--curvature coupling. In particular, we show how the presence of intrinsic spin leads to an additional contribution to the detector response beyond the conventional geodesic response. Within the pole--dipole approximation and adopting the Tulczyjew spin supplementary condition \cite{Tulczyjew}, we employ the worldline deviation equations for spinning particles developed in Ref.~\cite{Bini_2017} to derive the corresponding generalised detector response. This provides a direct way to distinguish/handle the spin-dependent response from the conventional geodesic response at the level of the measured strain. We illustrate this response for a monochromatic plane GW and estimate the magnitude of the spin-dependent correction for representative parameters relevant to present detectors. Towards the end,  we study the memory effect for spinning particles and explore  whether spin can contribute beyond conventional displacement memory. 

\noindent The remainder of this paper is organised as follows. In Sec. (\ref{secII}), we review the equations of motion for spinning test particles in the pole--dipole approximation. In Sec. (\ref{secIII}), we use the worldline deviation equation to investigate spin--curvature coupling in the weak-field metric and derive the corresponding modification to the detector strain. We illustrate the response using a tilted Michelson interferometer and a monochromatic GW, and provide representative order-of-magnitude estimates. In Sec. (\ref{secIV}), we analyze the memory effect associated with spinning test particles. Finally, Sec. (\ref{secV}) summarizes our results, discusses their physical implications and suggests possible future directions.

\section{Mathisson-Papapetrou-Dixon (MPD) equations}\label{secII}
\noindent The dynamics of an extended body in a prescribed gravitational background can be described through the multipole expansion of its energy--momentum tensor \cite{10.1098/rspa.1951.0200,10.1098/rspa.1970.0020}. In this formalism, the body is assumed to be a narrow world tube in spacetime, reflecting its finite spatial extent during evolution. A representative timelike worldline, lying inside this world tube, is introduced to describe the effective center-of-mass motion of the body. If $X^\mu$ denotes a point on this reference worldline and $x^\mu$ an arbitrary point within the body, the relative displacement is defined by $\delta x^\mu = X^\mu - x^\mu$.
The internal structure of the body is characterized by the moments of the energy--momentum tensor $T^{\mu\nu}$ about the reference worldline. These moments are obtained by integrating $T^{\mu\nu}$ weighted by increasing powers of the displacement vector ($\delta x^{\mu}$),
\begin{equation}
\int T^{\mu\nu}dV,\qquad
\int \delta x^\lambda T^{\mu\nu}dV,\qquad
\int \delta x^\lambda\delta x^\rho T^{\mu\nu}dV\ \ldots,
\end{equation}
where the integration is performed over a hypersurface of constant time. Physically, these integrals correspond to the monopole, dipole, quadrupole, and higher multipole moments of the body's energy--momentum distribution. Keeping only the monopole moment yields the single-pole approximation, in which the body is treated as a structureless test particle moving solely under the influence of the background geometry. Retaining, in addition, the dipole moment leads to the pole--dipole approximation, where the body's internal structure is represented entirely by its intrinsic spin while higher multipole effects, such as tidal deformations and finite-size corrections, are neglected. Higher-order approximations can be constructed systematically by including the quadrupole and subsequent multipole moments.

\noindent The pole--dipole approximation is valid under the assumption that the characteristic size of the body is much smaller than the typical curvature scale of the background spacetime \cite{10.1098/rspa.1970.0020}. In this regime, the higher multipole moments contribute only subdominant corrections to the dynamics, allowing the body's motion to be described accurately by its mass and intrinsic spin alone. In the formal limit
\begin{equation}
|\delta x^\mu|\rightarrow0,
\end{equation}
the body approaches an idealized point particle, and the multipole expansion becomes exact. Within the pole--dipole approximation, the intrinsic angular momentum of the body is described by the antisymmetric spin tensor \cite{Dixon:1964cjb, Dixon:1970zz}
\begin{equation}
S^{\mu\nu}= \int
\left(\delta x^\mu T^{0\nu} - \delta x^\nu T^{0\mu}\right)dV,
\end{equation}
which represents the first moment of the energy--momentum distribution. The antisymmetry of $S^{\mu\nu}$ indicates that the intrinsic angular momentum is associated with infinitesimal rotations about the center-of-mass worldline. Together with four-momentum, the spin tensor completely characterizes the dynamics of a pole--dipole particle governed by the MPD equations.
It also appears in the leading-order coupling between
the internal angular momentum of the body and the background
spacetime curvature, and forms the basis for the MPD equations governing the motion of spinning test particles \cite{Dixon:1964cjb, Dixon:1970zz, Mathisson:1937zz, Corinaldesi:1951pb, 10.1098/rspa.1951.0200, Steinhoff:2009tk}.
The MPD equations of motion are eventually  obtained by applying Einstein's field equations, together with the conservation of the energy-momentum tensor, $\nabla_{\mu}T^{\mu\nu}=0$, describing the body. For a single-pole, this leads to a free particle moving along the
geodesics associated with $g_{\mu\nu}$. Using the above procedure, two sets of equations
can be derived for the motion of a pole-dipole particle and its spin. These are:
\begin{eqnarray}
{\dot{P}}^{\mu}=
-\frac{1}{2}R^{\mu}_{\hspace{2mm}\nu\lambda\rho}V^{\nu}
S^{\lambda\rho}\label{MPD_eq1} , \\
{\dot{S}}^{\mu\nu}= P^{\mu}V^{\nu} - P^{\nu}V^{\mu} \label{MPD_eq2}.
\end{eqnarray}
Here, an overdot denotes the covariant derivative with respect to the affine parameter ($\tau$), which may be expanded in terms of the Christoffel symbols using the standard definition. Further, $V^{\mu}=\frac{dX^{\mu}(\tau)}{d\tau}$ is the timelike four-velocity of the particle, $R^{\mu}_{\nu \lambda \rho}$ is the
curvature tensor, $P^{\mu}=\int{T^ {0\mu}dV}$ represents the
particle's four-momentum and, more explicitly, the right-hand-side (R.H.S.) of  
 Eq. \eqref{MPD_eq1} expresses the coupling of the spin with curvature.
 We get back ordinary geodesic motion for a spinless ($S^{\mu \nu}=0$) particle. Unlike special relativity, $P^{\mu}$ and $V^{\mu}$ are generally not proportional to each other. 
 Eqs. \eqref{MPD_eq1} and \eqref{MPD_eq2} alone are not sufficient to yield a solution since the number of equations is less than the number of variables
 in the system. To remedy this
deficiency, different supplementary conditions have been proposed. Here, we adopt the following condition \cite{Costa:2014nta, Costa:2011zn}
\begin{equation}
P_{\mu}S^{\mu\nu}=0. \label{TD_SSC}
\end{equation}
If we take a covariant derivative of the above equation, one can deduce the following conserved quantities,
$
P^{\mu}P_{\mu}=\mbox{const.}=-m^2$, and $\frac{1}{2}S^{\mu\nu}S_{\mu\nu}=\mbox{const.}=S^2.$
One can express the particle's spin by the four-vector
\begin{eqnarray}
S^{\mu}{\equiv}\frac{1}{2m}{\eta}^{\mu\nu\lambda\kappa}P_{\kappa}
S_{\nu\lambda},
\end{eqnarray}
where ${\eta}^{\mu\nu\lambda\kappa}=\frac{1}{\sqrt{-g}}{\epsilon}^
{\mu\nu\lambda\kappa}$ is the alternating tensor with ${\epsilon}^{0123}=1$
and $g$ is the determinant of the metric. It then follows that
$P_{\mu} S^{\mu}=0$ and $S^{\mu}S_{\mu}=S^2$.
To determine the particle’s trajectory, one must first obtain its four-velocity. However, Dixon’s equations do not provide the four-velocity explicitly. Therefore, it is determined indirectly. One such
relation is \cite{Tod:1976ud, Mohseni:2000re}
\begin{equation}
(P\cdot P)V^{\mu}=(P\cdot V)\left( P^{\mu}-\frac{2R_{\sigma\nu\lambda\rho}
P^{\nu}S^{\mu\sigma}
S^{\lambda\rho}}{4P\cdot P-R_{\mu\nu\lambda\kappa}S^{\mu\nu}
S^{\lambda\kappa}}\right)
\end{equation}
Unlike for $P^{\mu}$, there is no guarantee from the equations of motion that $V^{\mu}$ remains timelike. As Dixon's equations are reparametrization invariant, it is
convenient to choose $
P^{\mu} V_{\mu}= -m$ \cite{Ehlers1977, Witzany:2018ahb}.
This is the gauge in which the instantaneous zero-momentum and zero-velocity frames are simultaneous \cite{Dixon:1970zz, Mohseni:2000re}.

\noindent With this setup, we next turn to the weak plane GW spacetime, applying the MPD equations to investigate how the spin–curvature coupling modifies the motion of spinning particles.

\section{Effect of a spinning particle in a plane GW spacetime}\label{secIII}
\noindent Let us now consider the weak plane GW spacetime, defined by the line element,
\begin{equation}
ds^2=-dt^2+dx^2+(1-h_{+})dy^2+(1+h_{+})dz^2-2h_{\times}dydz,
\label{weak_field_metric}
\end{equation}
where $h_{+}$ and $h_{\times}$ denote the waveforms of the two physical polarization modes of the GW. Here, the wave propagates along the positive $x$-direction. The retarded null coordinate is defined as,
\begin{equation}
u=t-x.
\end{equation}
In the transverse-traceless (TT) gauge, the metric perturbations depend only on the retarded time, so that $h_{+}=h_{+}(u)$, and $h_{\times}=h_{\times}(u)$. In the weak-field regime, $|h_{+}|,|h_{\times}|\ll1$, and we retain only linear terms in the metric perturbations. In this approximation, the background geometry remains flat while the GW degrees of freedom are entirely encoded in the transverse-traceless perturbations $h_{+}$ and $h_{\times}$.
The linearized Riemann tensor is given by
\begin{equation}
    R_{\alpha \beta \gamma \delta}= \frac{1}{2} \left(h_{\alpha \delta , \beta \gamma} + h_{\beta \gamma, \alpha \delta} - h_{\alpha \gamma , \beta \delta}- h_{\beta \delta , \alpha \gamma}\right).
\end{equation}
For the spacetime defined by Eq.~\eqref{weak_field_metric}, the nonzero linearized Riemann tensor components are: 
\begin{equation}
\begin{aligned}
  R_{0y0y}&=  \frac{1}{2} \ddot{h}_{+} &  R_{0z0z}&= - \frac{1}{2} \ddot{h}_{+} & R_{xyxy}&=  \frac{1}{2} \ddot{h}_{+} &  R_{xzxz}&= - \frac{1}{2} \ddot{h}_{+}, \\
   R_{0yxy}&= - \frac{1}{2} \ddot{h}_{+} &  R_{0zxz}&=  \frac{1}{2} \ddot{h}_{+} &  R_{0y0z}&=  \frac{1}{2} \ddot{h}_{\times}  & R_{xyxz}&=  \frac{1}{2} \ddot{h}_{\times}, \\
   R_{0yxz}&= - \frac{1}{2} \ddot{h}_{\times} & R_{0zxy}&= - \frac{1}{2} \ddot{h}_{\times},
\end{aligned}
\end{equation}
where $\dot{h}_{+,\times}\equiv \frac{\partial h_{+,\times}}{\partial u}$ and $\ddot{h}_{+,\times}\equiv \frac{\partial^2h_{+,\times}}{\partial u^2}$. These generate all remaining components through the Riemann tensor symmetries and the Bianchi identity.

\noindent For a spinning particle, using the MPD equations, the generalized deviation equation can be written as (for the detailed discussion see \cite{Bini_2017}), 
\begin{equation}
    \frac{D^2 Y^{\mu}}{d \tau^2}=- R^{\mu}_{\alpha \beta \gamma} U^{\alpha} Y^{\beta} U^{\gamma} + Y^{\rho} \nabla_{\rho} a^{\mu}.
    \label{worldline_deviation_equation}
\end{equation}
Here, $Y^{\mu}$ is the deviation vector, $U^{\alpha}$ is the 4-velocity and $\tau$ represent the proper time. The first term on the R.H.S. of the above equation corresponds to the ordinary geodesic motion, whereas the second term represents the MPD force. Here, $a^{\mu}$ is the MPD acceleration, defined as,
\begin{eqnarray}
    a^{\mu}= - \frac{1}{2m} R^{\mu}_{\nu \alpha \beta} U^{\nu} S^{\alpha \beta},
    \label{MPD_accelaration}
\end{eqnarray}
where $S^{\alpha \beta}$ is the antisymmetric spin tensor. It is clear that if the particle does not have any internal spin degree of freedom, the second term does not survive and we get back the ordinary geodesic deviation equation. Note
that the second term in Eq. \eqref{worldline_deviation_equation} contains the covariant derivative of \eqref{MPD_accelaration}, which acts on $R_{\nu \alpha \beta}^{\mu}$, $U^{\nu}$, and $S^{\alpha \beta}$. However, considering upto linear order in $h$ and $S$, it can be shown that only the term 
involving the covariant derivative of the Riemann tensor survives. To see this
explicitly, we note that (to linear order in spin) \cite{Toshmatov_2020}, 
\begin{equation}
P^\mu \sim m U^\mu  \implies \frac{DS^{\mu\nu}}{D\tau}=0. \label{Linear_S}
\end{equation}
Eq. \eqref{Linear_S} implies that the spin tensor is parallel transported along the fiducial worldline to linear order. Thus, the contribution of the covariant derivative of $S^{\alpha \beta}$ can be neglected in the linear order while solving the deviation equation. Now we study the contribution of the covariant derivative of $U^{\nu}$, which can be checked from Eq. \eqref{MPD_eq1}. In the weak plane GW spacetime, the linearized Riemann tensor satisfies $R^{\mu}_{\nu\alpha\beta}\sim\mathcal O(h)$. 
Expanding the momentum and four-velocity about the flat-spacetime solution, $P^\mu=P_0^\mu+\delta P^\mu$, $S^{\alpha \beta}= S_0^{\alpha \beta}+ \delta S^{\alpha \beta}$, and $U^\mu=U_0^\mu+\delta U^\mu$, the first-order MPD equation becomes
\begin{equation}
\frac{D(\delta P^\mu)}{D\tau}= -\frac{1}{2}
R^{\mu}_{\nu\alpha\beta}
U_0^\nu
S_0^{\alpha\beta},
\end{equation}
whose R.H.S. is of order $\mathcal O(hS)$. Assuming that the GW pulse has a finite duration, integration preserves the perturbative order, yielding $\delta P^\mu= \mathcal O(hS)$. Using $P^\mu \sim mU^\mu$, one immediately obtains $\delta U^\mu \sim \mathcal O(hS)$.
Consequently,
\begin{equation}
 R^{\mu}_{\nu\alpha\beta}U^\nu S^{\alpha\beta}= R^{\mu}_{\nu\alpha\beta}U_0^\nu S^{\alpha\beta}+ R^{\mu}_{\nu\alpha\beta}\delta U^\nu S^{\alpha\beta},
\end{equation}
where the second term is of higher perturbative order $\mathcal O(h^2 S^2)$. As a result, we can neglect the second term. Thus the deviation equation becomes,
\begin{equation}
    \frac{D^2 Y^{\mu}}{d \tau^2}=- R^{\mu}_{\alpha \beta \gamma} U^{\alpha} Y^{\beta} U^{\gamma} - \frac{1}{2m} Y^{\rho} \left(\nabla_{\rho}R^{\mu}_{\nu \alpha \beta}\right) U^{\nu} S^{\alpha \beta}.
    \label{final_worldline_deviation_equation}
\end{equation}Therefore, to linear order in the weak-field and spin expansion, the MPD acceleration and its gradient may be evaluated on the background fiducial worldline with $U^\mu=U_0^\mu=(1,0,0,0)$,
while terms involving $\delta U^\mu$ or its covariant derivative are consistently neglected.

\noindent Further, the spin degrees of freedom are fixed by imposing the Tulczyjew-Dixon spin supplementary condition (SSC) \cite{Tulczyjew},
\begin{equation}
  S^{\mu\nu}P_\nu=0
\quad\Longrightarrow\quad
S^{\mu\nu}U_\nu=0.  
\end{equation}
Using $U^\mu=(1,0,0,0)$, one gets $S^{0 \mu}=0$. Only the spatial components ($S^{12}, S^{13}, S^{23})$ remain. 
Thus, to the order of interest ($\mathcal O(hS)$), the fiducial MPD observer is completely characterized by a static four-velocity and a constant spatial spin vector. With this setup established for the worldline deviation equation and the corresponding spin configuration, we next carry out the strain analysis in the weak plane GW spacetime including the spin contribution, and derive the spin-induced effects on memory.


\subsection{GW strain analysis}
\noindent We begin by writing the spin antisymmetric tensor as 
\begin{eqnarray}
    S^{ij}= \epsilon^{ijk} S_k.
\end{eqnarray}
The MPD acceleration term along the chosen fiducial observer ($U^{\mu}= (1, 0, 0, 0)$) becomes,
\begin{equation}
    a^i= -\frac{1}{2m} R_{0jk}^i \epsilon^{jkl} S_{l}.
\end{equation}
As a result, the MPD acceleration gradient term reduces to
\begin{equation}
    Y^{\rho} \nabla_{\rho} a^{i}= Y^{\rho} \partial_{\rho} a^i= -\frac{1}{2m} \left(Y^0- Y^1\right)\partial_0(R_{m0jk}) \eta^{im}\epsilon^{jkl} S_l.
\end{equation}
Using the linearised Riemann tensor definition, we find 
$R_{m0jk}= \frac{1}{2} \left(\partial_0 \partial_j h_{mk}- \partial_0 \partial_k h_{mj}\right)$. Putting all of the above together, we can rewrite the deviation equation as, 
\begin{equation}
    \ddot{Y}_{\alpha}= \frac{1}{2} \ddot{h}_{\alpha \beta} Y^{\beta}-\frac{(Y^0-Y^1)}{4m} S_l \left(\partial_j \ddot{h}_{\alpha k}- \partial_k \ddot{h}_{\alpha j}\right)\epsilon^{jkl}.
    \label{devi_eq_no_contraction}
\end{equation}
As mentioned before, throughout our work, an overdot denotes differentiation $\frac{\partial}{\partial u}\equiv \frac{\partial}{\partial t}\equiv \frac{\partial}{\partial\tau}$ given the worldline $x=x_0=constant$ with $u=t-x_0$, when evaluated on the fiducial observer's trajectory.
The temporal and longitudinal components of the deviation vector satisfy the equations
\begin{equation}
\ddot{Y}_0=0,
\qquad
\ddot{Y}_1=0,
\end{equation}
whose general solutions are
\begin{equation}
Y_0(t)=at+b,
\qquad
Y_1(t)=ct+d,
\end{equation}
where $a,b,c$, and $d$ are integration constants determined by the initial conditions.  We define, 
\begin{equation}
    Y^u\equiv Y^0-Y^1,
    \label{def_y_u}
\end{equation} which becomes $Y^u=(a-c)t+(b-d)$. However, the linear term vanishes when $\dot{Y}_0=\dot{Y}_1$, which essentially implies $a=c$. Thus, $Y^u=b-d$ is a constant.
Throughout this work, we impose this condition and, consequently, denote the constant background value simply by  $Y^u$. Physically, since $\ddot{Y}^0=\ddot{Y}^1=0$ identically, both before and after the wave passes as well as during its passage, $Y^u$ is not generated by the GW. Instead, it characterizes the initial detector configuration and enters Eq.~\eqref{devi_eq_no_contraction} as a constant coefficient of the spin–curvature force term. Further, using the perturbative expansion ($Y^i= Y_0^i+ \delta Y^i$ and $S_l= S^0_l+ \delta S_l$) and keeping up to the linear order term, we get, 
\begin{equation}
    \delta \ddot{Y}_{\alpha}= \frac{1}{2} \ddot{h}_{\alpha \beta} Y_0^{\beta}- \frac{Y^u}{4m} S_l^0 (-2 \dddot{h_{\alpha k}}) \epsilon^{1kl}.
\end{equation}
To get the above from Eq. \eqref{devi_eq_no_contraction}, we have used the fact that the GW is propagating along the $x$ direction, and the waveform
will only depend on the ($t, x$) coordinates. Also, $\partial_0= -\partial_1$ from $u=t-x$. Thus, the change in the components of the deviation vector becomes 
\begin{equation}
    \delta \ddot{Y}_{\alpha}= \frac{1}{2} \ddot{h}_{\alpha \beta} Y_0^{\beta}+ \frac{Y^u}{2m} S_l^0 \dddot{h}_{\alpha k}\epsilon^{1kl}.
\end{equation}
Integrating twice, we get,
\begin{equation}
    \delta Y_{\alpha}= \frac{1}{2} h_{\alpha \beta} Y_0^{\beta} + \frac{Y^u}{2m} \dot{h}_{\alpha k} \epsilon^{1kl} S_l^0.
\end{equation}
To determine the GW response of an interferometric detector, one must project the change in the deviation vector onto the detector frame. Let $n_{(1)}^\alpha$ and $n_{(2)}^\alpha$ denote the unit vectors along the two interferometer arms. Accordingly, the projected components along the two detector arms are given by
\begin{eqnarray}
    \delta Y_1= n^{\alpha}_{(1)} \delta Y_{\alpha}= \frac{1}{2} h_{\alpha \beta} L n^{\alpha}_{(1)} n^{\beta}_{(1)} + \frac{Y^u}{2m} \dot{h}_{\alpha k} \epsilon^{1kl} S_{(0)} n_{l (1)}n^{\alpha}_{(1)},  \\
    \delta Y_2= n^{\alpha}_{(2)} \delta Y_{\alpha}= \frac{1}{2} h_{\alpha \beta} L n^{\alpha}_{(2)} n^{\beta}_{(2)} + \frac{Y^u}{2m} \dot{h}_{\alpha k} \epsilon^{1kl} S_{(0)} n_{l (2)}n^{\alpha}_{(2)},
\end{eqnarray}
where we have considered the initial arm-lengths to be $L$, and thus we can write $Y^{\beta}_0= L n^{\beta}$. As we are in the TT gauge, the wave frame and the detector frame are basically the same. Now one can write $h_{\alpha \beta}$ and its derivative in terms of the polarization basis as $h_{\alpha \beta}= h_{+} e_{\alpha \beta}^{+} + h_{\times} e_{\alpha \beta}^{\times}$. Consequently, $\dot{h}_{\alpha \beta}= \dot{h}_{+} e_{\alpha \beta}^{+} + \dot{h}_{\times} e_{\alpha \beta}^{\times}$. The detector strain is defined as the differential fractional change in the arm lengths, obtained from the projected changes in the deviation vector. Thus, the strain can be written as 
\begin{equation}
\begin{split}
    h_D= \frac{\delta Y_1 - \delta Y_2}{L}= \frac{1}{2} \left(h_{+} e_{\alpha \beta}^{+} +h_{\times} e_{\alpha \beta}^{\times}\right) \left(n^{\alpha}_{(1)} n^{\beta}_{(1)}-n^{\alpha}_{(2)} n^{\beta}_{(2)}\right), \\ + \frac{Y^u S_{(0)}}{2 L m} \left(n_{l (1)} n^{\alpha}_{(1)}- n_{l (2)} n^{\alpha}_{(2)}\right) \epsilon^{1kl} \left(\dot{h}_{+} e_{\alpha k}^{+} + \dot{h}_{\times} e_{\alpha k}^{\times}\right).
\end{split}\label{strainfinal}
\end{equation}
The detector tensor is defined as $D^{\alpha \beta}=\frac{1}{2} (n^{\alpha}_{(1)}n^{\beta}_{(1)}- n^{\alpha}_{(2)} n^{\beta}_{(2)})$, and the antenna pattern functions are defined as $F_{+}= D^{\alpha \beta} e_{\alpha \beta}^{+}$ and $F_{\times}= D^{\alpha \beta} e_{\alpha \beta}^{\times}$. Also, $\epsilon^{1kl} e_{\alpha k}^{+}= e_{\alpha l}^{\times}$ and $\epsilon^{1 k l} e_{\alpha k}^{\times}= - e_{\alpha l}^{+}$ \cite{Maggiore:2007ulw, Creighton:2011zz}.
Hence, using the above definitions, one can rewrite the strain as 
\begin{equation}
   h_D= \left(h_{+}- \frac{Y^u S_{(0)} }{L m} \dot{h}_{\times}\right) F_{+} + \left(h_{\times}+ \frac{Y^u S_{(0)} }{L m} \dot{h}_{+}\right) F_{\times}.
    \label{final_GW_strain}
\end{equation}
The important features in the above analysis and expression are as follows:
\begin{itemize}

\item  The additions in the amplitudes along $F_+$ and $F_\times$ are proportional to the time derivatives of the orthogonal polarization modes, leading to a mixing between the original plus and cross polarizations which is absent for spinless test masses. In the spinless limit, $S_{(0)}\rightarrow 0$, the standard detector response $h_D = F_{+}h_{+} + F_{\times}h_{\times}$ is recovered exactly.

\item  The spin-induced contribution does modify the standard antenna pattern functions ($F_{+}, F_{\times}$) through a tilt of the
interferometer arms manifest in the new definition of the normals $n^\alpha_{(1)}$ 
and $n^\alpha_{(2)}$, described in the subsection \ref{subsection3.2}. The detector's directional sensitivity is affected by this tilt, while the net contributions along the new $F_+$ and $F_\times$ are modified by effects originating in the spin-curvature coupling.

\end{itemize}
While Eq. (\ref{final_GW_strain}) has been derived for a general detector configuration; however, its physical understanding naturally raises the question of whether such a detector geometry can be realized while preserving the fundamental properties of an interferometric detector, which we attempt to address below.

\subsection{Geometrical realization of a tilted interferometric detector}\label{subsection3.2}

\noindent The generalized detector response obtained in the previous subsection depends explicitly on $Y^u$, which characterizes the separation of neighbouring spinning test objects along the propagation direction of the GW. Conventional Michelson interferometers are purely transverse and therefore satisfy $Y^u=0$, for which the spin-dependent correction vanishes identically \cite{Saulson:2017jlf, Maggiore:2007ulw, Dhurandhar:2022eik}. So it is natural to ask whether a detector configuration admitting a finite, nonzero $Y^u$ can be constructed while preserving the orthogonality of the interferometer arms. In this subsection, we attempt to demonstrate that such a configuration can be realized through a rigid rotation of the $Y^u=0$ detector plane. This construction includes a non-zero parameter $Y^u$ appearing in the detector response and establishes how the detector tensor and antenna pattern functions are modified under such a rotation.

\noindent The rigid rotation preserves the orthogonality and equal arm lengths of the Michelson interferometer while introducing a non-vanishing longitudinal component into the detector geometry. Consequently, the detector tensor and antenna pattern functions capture an explicit tilt angle ($\alpha$)-dependence, which modifies the directional response of the interferometer even within the standard geodesic case. In the absence of intrinsic spin, one may consider a tilted detector configuration, doing so does not lead to any additional physical effect, since the detector response reduces exactly to the conventional geodesic response for the corresponding detector orientation. As a result, in the spinless case, the introduction of a finite longitudinal component is not required/necessary. The physical significance of the tilted geometry becomes important only when the intrinsic spin is present, where the non-vanishing $Y^u$ enters directly through the spin-curvature coupling. Thus, the finite longitudinal separation provided by the tilted detector gives the necessary geometrical ingredient for the spin-dependent correction derived in Eq. (\ref{final_GW_strain}).

\noindent Note that the purpose of this construction is not to propose an immediately realizable detector, but to demonstrate that the non-zero longitudinal separation required by the spin-dependent detector response is geometrically consistent within an idealized Michelson interferometer. It therefore provides a proof of principle for future high-precision novel detector concepts capable of probing spin–curvature effects.

\begin{figure}[h]
\centering
\includegraphics[width=0.87
\textwidth]{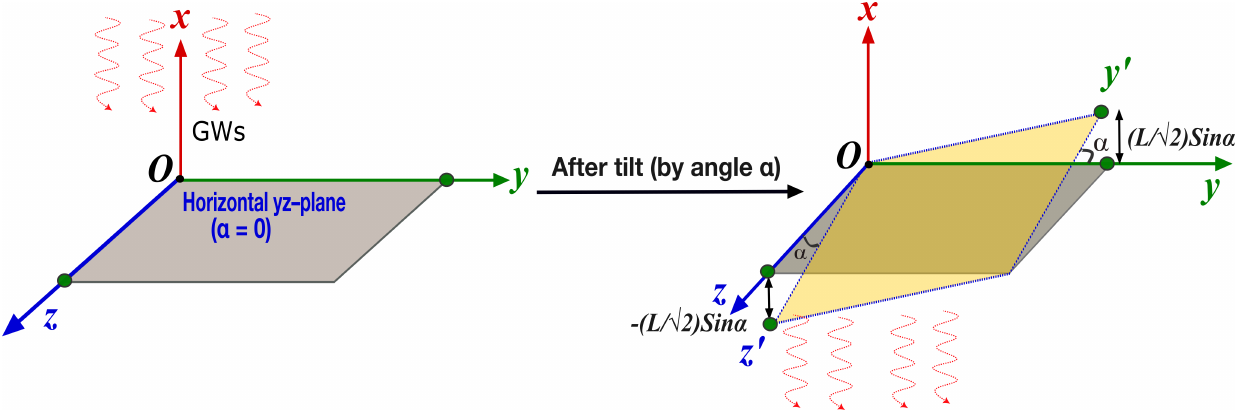}
\caption{Schematic illustration of a Michelson interferometer with a tilted detector frame. The original $yz$ plane and the tilted $y^\prime z^\prime$ plane (yellow) are inclined by an angle $\alpha$, with the corresponding arm directions indicated by $y$, $z$ and $y^\prime$, $z^\prime$.}
\label{schematic}
\end{figure}

\noindent To construct such a detector, as indicated in Fig. (\ref{schematic}), we begin with a standard Michelson interferometer which consists of a beam splitter located at the origin and two orthogonal arms of equal length $L$. In the unrotated configuration (tilt-angle $\alpha=0$), the unit vectors associated with the end test masses are located at (${n}_{(1)}, {n}_{(2)}$). In such a conventional Michelson interferometer, the detector plane lies entirely in the transverse $yz$-plane. To generate a finite longitudinal component and preserve the symmetry of the detector, the interferometer must be rotated as a rigid body. The tilt/rotation is chosen such that the two test masses acquire, in general, different longitudinal displacements, while preserving the equal arm lengths and mutual orthogonality of the Michelson interferometer. Once the detector plane is tilted, the wave frame and detector frame are no longer coincident, as they were in the untilted configuration. Therefore, the beam splitter must be tilted together with the detector arms so that the laser beams remain aligned with the test masses. The resulting signal measured in this tilted detector frame is then related to the theoretical response through the corresponding rotation of the original detector plane, implemented by the effective wave-to-detector transformation $M_{\text{eff}}$, as will be shown below. A rotation about either detector arm would tilt only one arm and therefore break the symmetry between the two interferometer arms. Instead, we rotate the detector about the axis
\begin{equation}
\hat{\mathbf a}
=
\frac{1}{\sqrt2}
(\hat{\mathbf y}+\hat{\mathbf z}),
\label{rotationaxis}
\end{equation}
which bisects the angle between the two arms and treats them on an equal footing. The rigid rotation through an angle $\alpha$ is conveniently described by Rodrigues' rotation formula \cite{Goldstein2002, Murray1994},
\begin{equation}
\mathbb{R}(\alpha) = \mathbb{I}+\mathbb{K}\sin\alpha + (1-\cos\alpha)\mathbb{K}^2,
\label{Rodrigues}
\end{equation}
where $\mathbb{K}$ denotes the skew-symmetric matrix associated with the rotation axis $\hat{\mathbf a}$. The explicit form of the rotation matrix is given in Appendix~(\ref{app:A}). The rotated/tilted detector arms are then obtained as
\begin{equation}
n'_{(i)} = \mathbb{R}(\alpha)\,
n_{(i)}, \qquad i=1,2.
\label{rotatedarms}
\end{equation}
with $n'_{(1)} = \Big\{-\frac{\sin{\alpha}}{\sqrt{2}}, \frac{1}{2}(1+\cos\alpha), \frac{1}{2}(1-\cos\alpha)\Big\}$ and $n'_{(2)} = \Big\{\frac{\sin{\alpha}}{\sqrt{2}}, \frac{1}{2}(1-\cos\alpha), \frac{1}{2}(1+\cos\alpha)\Big\}$. Since $\mathbb{R}(\alpha)$ is an orthogonal rotation matrix satisfying $\mathbb{R}^{\text{T}}\mathbb{R}=\mathbb{I}$, the rotated arm vectors preserve their unit norms and mutual orthogonality, i.e., $n'_{i}n'_{j}=\delta_{ij}$. Here, prime denotes the quantities associated with the tilted plane. The Eq. (\ref{rotatedarms}) can also be written equivalently in indexed notation as: $n_{(i)}^{\prime\,\alpha} =
R^{\alpha}{}_{\beta}(\alpha)\,
n_{(i)}^{\beta}$. The $R^\alpha{}_\beta$ denotes the matrix components of the rotation operator $\mathbb{R}$. Additionally, since the GW propagates along the $x$-direction, the longitudinal quantity is identified with the $x$-component of the rotated initial separation vector. Using $Y_0^{\prime\alpha}=Ln_{(i)}^{\prime\alpha}$, we obtain the magnitude of the following form
\begin{equation}
|Y^u| = |Y_0^{\prime x}| =
L|n_{(i)}^{\prime x}| =
\frac{L}{\sqrt2}\sin\alpha,
\label{Yu_alpha}
\end{equation}
where the last equality follows from the explicit form of the rotated arm vectors.

\noindent With this, the detector tensor associated with the tilted interferometer is constructed from the rotated arm vectors as
\begin{equation}
D^{\prime\,\alpha\beta}
=\frac{1}{2}
(n_{(1)}^{\prime\,\alpha}
n_{(1)}^{\prime\,\beta}
-
n_{(2)}^{\prime\,\alpha}
n_{(2)}^{\prime\,\beta}).
\label{rotated_detector_tensor}
\end{equation}
Using Eq.~(\ref{rotatedarms}), the detector tensor can equivalently be expressed as
\begin{equation}
D^{\prime\,\alpha\beta}
=
R^{\alpha}{}_{\mu}
R^{\beta}{}_{\nu}
D^{\mu\nu},
\label{detector_tensor_transformation}
\end{equation}
where $D^{\mu\nu} =\frac{1}{2}( n_{(1)}^{\mu}n_{(1)}^{\nu} - n_{(2)}^{\mu}n_{(2)}^{\nu})$
is the detector tensor in the untilted configuration, given around Eq. (\ref{strainfinal}). The rotation modifies the relative orientation between the detector and wave frames. Consequently, the standard/usual wave-to-detector transformation matrix $\mathbb{M}$ (Euler matrix) is replaced by the effective transformation
\begin{equation}
\mathbb{M}_{\rm eff}
=
\mathbb{M}(\Phi,\Theta,\Psi)\,
\mathbb{R}^{\rm T}(\alpha),
\label{Meff}
\end{equation}
with the corresponding indexed form $M_{\rm eff}^{A}{}_{\beta} =
M^{A}{}_{\alpha}
(R^{\rm T})^{\alpha}{}_{\beta}$. This determines the effective transformation between the wave and detector frames used in evaluating the antenna pattern functions in the tilted configuration. Further, using the rotated detector tensor Eq.~(\ref{rotated_detector_tensor}) together with the effective transformation matrix (\ref{Meff}), the polarization tensors $e^{+,\times}_{\alpha\beta}$ are projected onto the tilted detector frame. The corresponding antenna pattern functions are given by
\begin{equation}
F'_+
=
D^{\prime\alpha\beta}e^{+}_{\alpha\beta},
\qquad
F'_\times
=
D^{\prime\alpha\beta}e^{\times}_{\alpha\beta}.
\label{tilted_antenna}
\end{equation}
To reiterate, the rotated detector tensor $D'^{\alpha\beta}$ is defined in the tilted detector frame, whereas the polarization tensors are defined in the wave frame; the two frames are related through the effective transformation matrix ~(\ref{Meff}). As a result, the antenna pattern functions carry an explicit dependence on both the Euler angles $(\Phi,\Theta,\Psi)$ \cite{Goldstein2002, Dhurandhar:2022eik}, which specify the relative orientation between the source and detector, and the detector tilt angle $\alpha$, which characterises the tilted interferometer geometry. Thus,
\begin{align}
F'_{+,\times}\equiv F'_{+,\times}(\Phi,\Theta,\Psi;\alpha). \label{tiltedantenna}
\end{align}
Their complete analytical expressions and related details are presented in Appendix~(\ref{app:A}). Thus, the tilted antenna pattern functions provide the geometric realisation required by the generalised detector response. This establishes a consistent physical interpretation of the underlying setup and identifies detector orientation as an additional geometrical handle through which spin--curvature effects may, in principle, be probed in future high-precision GW experiments.

\subsection{Monochromatic wave example}
\noindent To illustrate the physical implications of the generalized detector response, we consider a monochromatic plane GW propagating along the $x$$-$direction. The two independent transverse--traceless polarization modes are assumed to be periodic functions of the retarded time $u=t-x$, and are given by
\begin{equation}
    h_{+}(u)=A\cos(\omega u), \qquad
    h_{\times}(u)=B\sin(\omega u),
\end{equation}
where $A$ and $B$ denote the amplitudes of the plus and cross polarizations, respectively, and $\omega$ is the angular frequency of the incoming GW. 
Substituting these expressions into the generalized detector response given by Eq. \eqref{final_GW_strain},
we obtain
\begin{equation}
\begin{aligned}
h_D
=&
\left(A
-\frac{\omega Y^uS_{(0)}}{Lm} B\right)
F_{+}\cos(\omega u)
+\left(B
-\frac{\omega Y^uS_{(0)}}{Lm}A\right)
F_{\times}\sin(\omega u).
\end{aligned}
\label{eq:monochromatic_strain}
\end{equation}
Eq.~(\ref{eq:monochromatic_strain}) shows that the spin--curvature interaction modifies the effective amplitudes of the two polarization modes while preserving their harmonic time dependence. Defining the effective polarization amplitudes as
\begin{equation}
\tilde{A} = A-\frac{\omega Y^uS_{(0)}}{Lm}B, \qquad
\tilde{B} = B-\frac{\omega Y^uS_{(0)}}{Lm}A, \label{monoamplitude}
\end{equation}
the detector response can be written in the compact form
\begin{equation}
h_D= \tilde{A}F_{+}\cos(\omega u) +\tilde{B}F_{\times}\sin(\omega u). \label{monostrain}
\end{equation}

\noindent The above expression demonstrates that the spin of the test body induces a mixing between the plus and cross polarization amplitudes through the spin--curvature coupling. The amplitudes acquire linear spin-dependent corrections proportional to the parameter $\frac{\omega Y^uS_{(0)}}{Lm}$.
In the spinless limit $S_{(0)}\rightarrow0$, the effective amplitudes reduce to $\tilde{A}=A$ and $\tilde{B}=B$;
thereby recovering the standard detector response
\begin{equation}
h_D
=
F_{+}A\cos(\omega u)
+
F_{\times}B\sin(\omega u).
\end{equation}

\noindent The spin--curvature contribution is controlled by the parameter $\frac{\omega Y_0^uS_{(0)}}{Lm}$ and is therefore expected to be amplified for GW signals with significant high-frequency. Moreover, the effect is intrinsically tied to the existence of a non-vanishing longitudinal component of the deviation vector through $Y^u$. This suggests that, in addition to the conventional transverse arm separation, a detector must admit a finite and time-independent longitudinal deviation in order for the spin--curvature coupling to leave an observable imprint on the measured strain.

\noindent Further, to characterize the relative magnitude of the two contributions, we consider representative parameters for a ground-based GW detector such as LIGO. We take corresponding characteristic strain amplitudes $A\simeq B$, antenna-pattern functions to be of order unity and $|\sin{(\omega u)}|$, $|\cos{(\omega u)|}\leq 1$. This estimate is intended only to illustrate the characteristic scale of the effect, i.e., an order of magnitude estimate, rather than to provide a detector-specific sensitivity forecast. From Eq.~(\ref{final_GW_strain}), the detector strain can be separated schematically into the conventional geodesic ($h_{D}^{\rm geo}$) contribution and the spin-dependent correction ($h_{D}^{\rm spin}$),
\begin{equation}
h_D=h^{\rm geo}_{D}+h^{\rm spin}_{D},
\end{equation}
which can be seen in Eq. (\ref{eq:monochromatic_strain}) for our case. For the monochromatic waveform considered above, their characteristic amplitudes scale as $h_D^{\rm geo}\sim A$ and $h^{\rm spin}_{D}\sim
\frac{\omega A}{c^2} \frac{Y^u S_{(0)}}{mL}$. Here, we have restored the factor of speed of light $c^{-2}$ when expressing the result in SI units. As expected, unlike the conventional geodesic response, the spin-dependent correction depends explicitly on the longitudinal deviation $Y^{u}$ as well as on the enhanced GW frequency ($\omega$). With the representative LIGO values \cite{Lasky:2016knh, LIGOScientific:2016aoc, Mohseni:2000re}, the conventional strain is of order $10^{-21}$, whereas the spin-dependent contribution is parametrically much smaller, depending on the corresponding values of $Y^{u} S_{(0)}/m$. The resulting estimates are summarized in Table~\ref{tab:strain_estimates}. A corresponding estimate for LISA is beyond the scope of the present analysis, which is valid in the limit, $L \ll \lambda_{\text{GW}}$. For the frequencies considered here, this condition is not satisfied by LISA, whose arm length is
expected to be around $L \approx 2.5 \times 10^9$
m \cite{2002PhRvD..66l2001S}. 


\begin{table}[h]
\centering
\footnotesize
\caption{Representative order-of-magnitude estimates of the geodesic and spin-dependent contributions to the detector strain for a monochromatic GW.}
\label{tab:strain_estimates}
\begin{tabular}{lc}
\toprule
\textbf{Quantity} & \textbf{LIGO} \\
\midrule

Representative GW amplitude $(A\simeq B)$
& $\sim 10^{-21}$ \\

Arm length $L$
& $4\times10^{3}\,\mathrm{m}$ \\

Angular frequency $\omega$
& $\sim 10^{3}\,\mathrm{s^{-1}}$ \\

Geodesic strain amplitude ($h_{D}^{\rm geo}$)
& $\sim 10^{-21}$ \\

Spin-dependent strain amplitude ($h_{D}^{\rm spin}$)
& $\displaystyle
\sim 2.8\times10^{-39}
\frac{Y^u S_{(0)}}{m}
$ \\

\bottomrule
\end{tabular}
\end{table}
\noindent Thus, the spin-curvature coupling produces a frequency-dependent correction to the oscillatory detector response, but its magnitude is suppressed relative to the conventional geodesic contribution for the representative parameters considered here. \\
\noindent Similarly, one can do this analysis using the rotated unit vectors along the two arms of the tilted interferometer and constructing the associated detector tensors. Explicitly, this amounts to replacing $F_{+,\times}$ in Eq. \eqref{monostrain} by the tilted antenna pattern functions $F'_{+,\times}(\Phi,\Theta,\Psi;\alpha)$ of Eq. \eqref{tiltedantenna} and (see Appendix \ref{app:A}), while the amplitudes 
($\tilde{A}, \tilde{B}$) of Eq. \eqref{monoamplitude} retain their form with $Y^u$ fixed by the tilt through Eq. \eqref{Yu_alpha}. So the spin-dependent correction term will carry a dependence on the tilt angle $\alpha$ and vanish smoothly as $\alpha\rightarrow 0$, which is consistent with $Y^u=0$ for the conventional untilted Michelson configuration. Although the tilt modifies the numerical coefficient of the spin-dependent response through the tilted antenna-pattern functions and the corresponding $Y^u(\alpha)$ dependence, these angular/trigonometric factors remain bounded and of order unity and do not introduce any parametrically large changes. Therefore, the tilt does not change the order of magnitude of the spin-dependent response for the representative configurations considered here. Consequently, the estimates presented in Table \ref{tab:strain_estimates} remain unchanged at the level of order of magnitude.
\noindent Let us now compute the memory associated with the spin-dependent contribution by examining the asymptotic behaviour of the deviation and the change in deviation.

\section{Memory effect in plane GW spacetime: Spin contribution}\label{secIV}
\noindent We begin with the component form of Eq. \eqref{devi_eq_no_contraction}, from which the deviation equations are written as follows: 
\begin{eqnarray}
    \ddot{Y}^{2}=& -\frac{1}{2} \ddot{h}_{+} Y^2- \frac{1}{2} \ddot{h}_{\times} Y^3 -\frac{Y^u}{2m} \left[\dddot{h}_{+} S^{12} + \dddot{h}_{\times}S^{13}\right] \label{y_component_devi_eq},\\
    \ddot{Y}^{3}=& -\frac{1}{2} \ddot{h}_{\times}Y^2+\frac{1}{2} \ddot{h}_{+} Y^3 -\frac{Y^u}{2m} \left[\dddot{h}_{+} (- S^{13}) + \dddot{h}_{\times}S^{12}\right] \label{z_component_devi_eq}.
\end{eqnarray}

\noindent For the perturbative solution, we can write $Y^{A}= Y_{0}^{A} + \delta Y^{A}$ and $S^{\mu \nu}= S_0^{\mu \nu}+ \delta S^{\mu \nu}$. To the first order in the wave amplitude, $Y_{0}^{A}$ and $S_0^{\mu \nu}$ are constant. Hence, keeping only the terms up to $\mathcal{O} (hS)$, we get 
\begin{eqnarray}
    \delta \ddot{Y}^{2}=&-\frac{1}{2} \ddot{h}_{+} Y_0^2- \frac{1}{2} \ddot{h}_{\times} Y_0^3 -\frac{Y^u}{2m} \left[\dddot{h}_{+} S_0^{12} + \dddot{h}_{\times}S_0^{13}\right], \label{y_comp_devi_change}\\
    \delta \ddot{Y}^{3}=& -\frac{1}{2} \ddot{h}_{\times}Y_0^2+\frac{1}{2} \ddot{h}_{+} Y_0^3 -\frac{Y^u}{2m} \left[\dddot{h}_{+} (- S_0^{13}) + \dddot{h}_{\times}S_0^{12}\right]. \label{z_comp_devi_change}
\end{eqnarray}  
Integrating twice and considering $\delta Y^A (-\infty)= \delta \dot{Y}^A (-\infty)=0$, one can write the change in the deviation components as 
\begin{eqnarray}
    \delta Y^{2}=& -\frac{1}{2} h_{+} Y_0^2- \frac{1}{2} h_{\times} Y_0^3 -\frac{Y^u}{2m} \left[\dot{h}_{+} S_0^{12} + \dot{h}_{\times}S_0^{13}\right], \\
    \delta Y^{3} =& -\frac{1}{2} h_{\times}Y_0^2+\frac{1}{2} h_{+} Y_0^3 -\frac{Y^u}{2m} \left[\dot{h}_{+} (- S_0^{13}) + \dot{h_{\times}}S_0^{12}\right].
\end{eqnarray}
The GW memory effect is characterized by a permanent change in the relative motion of freely falling test particles following the passage of a GW. In the present analysis, $Y^A$ denotes the deviation vector and $\delta Y^A$ its change in the deviation induced by the GW, where $A=(y,z)$ denotes the transverse spatial directions. The memory observable is then defined by comparing the asymptotic changes in the relative motion between the far future and far past. Thus, the displacement and velocity memory are respectively
\begin{equation}
\Delta(\delta Y^A)
\equiv
\delta Y^A(+\infty)-\delta Y^A(-\infty),
\qquad
\Delta(\delta\dot{Y}^A)
\equiv
\delta\dot{Y}^A(+\infty)-\delta\dot{Y}^A(-\infty).
\end{equation}
A non-zero $\Delta(\delta Y^A)$ therefore quantifies a permanent change in the relative displacement, while a nonvanishing $\Delta(\delta\dot{Y}^A)$ corresponds to a permanent change in the relative velocity after the GW has passed.
The deviation equations given in Eqs. \eqref{y_comp_devi_change} and \eqref{z_comp_devi_change} can now be integrated successively to obtain the corresponding memory observables. Thus, we obtain
\begin{eqnarray}
 \Delta \delta\dot{Y}^y=&-\frac{1}{2} \left(\Delta \dot{h}_{+} Y^2_0 +\Delta \dot{h}_{\times} Y_0^3\right)- \frac{Y^u}{2m} \left[\Delta \ddot{h}_+ S_0^3-\Delta \ddot{h}_{\times} S_0^2\right], \label{y_component_vel_mem} \\
 \Delta \delta \dot{Y}^z=&-\frac{1}{2} \left(\Delta \dot{h}_{\times} Y^2_0 -\Delta \dot{h}_{+} Y_0^3\right)-\frac{Y^u}{2m} \left[\Delta \ddot{h}_+ S_0^2+ \ddot{h}_{\times} S_0^3\right], \label{z_componet_vel_mem}
\end{eqnarray}
which describes the velocity memory, while
\begin{eqnarray}
 \Delta \delta Y^y=&-\frac{1}{2} \left(\Delta h_{+} Y^2_0 +\Delta h_{\times} Y_0^3\right)-\frac{Y^u}{2m} \left[\Delta \dot{h}_+ S_0^3- \Delta \dot{h}_{\times} S_0^2\right] ,\label{y_component_displacement_mem} \\
    \Delta \delta Y^z=&-\frac{1}{2} \left(\Delta h_{\times} Y^2_0 -\Delta h_{+} Y_0^3\right)+ \frac{Y^u}{2m} \left[\Delta \dot{h}_+ S_0^2+ \Delta \dot{h}_{\times} S_0^3\right] ,\label{z_component_displacement_mem}
\end{eqnarray}
gives the corresponding displacement memory. ($\Delta h_{+,\times}, \Delta \dot{h}_{+,\times}$) are the difference between values in the asymptotic past and future null infinity. In the above expressions, the antisymmetric spin tensor has been expressed in terms of the spatial spin vector using the three-dimensional Levi-Civita tensor, so that the memory observables are written entirely in terms of the spin four vector. 

\noindent An important structural constraint follows from Eqs.~(\ref{y_component_displacement_mem}) and~(\ref{z_component_displacement_mem}). The ordinary geodesic contribution is determined by the change in the GW strain $\Delta h_{+,\times}$, whereas the spin-dependent contribution is determined by the change in its first derivative $\Delta\dot{h}_{+,\times}$. For physical GW signals from isolated sources, we consider bounded waveforms whose strain remains finite at early and late times; a memory signal may approach different finite values at the two ends. One can then show that the latter vanishes for such bounded waveforms, provided 
$\dot{h}_{\text{A}}(u)$ has a finite asymptotic limit as $u\to+\infty$;
$\lim_{u\to+\infty}\dot h_{\text{A}}(u)=c_{\text{A}}$, where $A\in (+,\times)$. At first sight, a nonzero $c_{\text{A}}$ might appear to provide a nonvanishing contribution to the spin-induced memory. However, boundedness of $h_{\text{A}}$ forces these asymptotic limits to vanish. To see this, using the fundamental theorem of calculus \cite{apostol1967calculus},
\begin{equation}
h_{\text{A}}(u)-h_{\text{A}}(u_0)
=
\int_{u_0}^{u}\dot h_{\text{A}}(\tilde u)\,d\tilde u.
\end{equation}
Note that $\dot{h}_{\text{A}}(u)$ approaches $c_{\text{A}}$ asymptotically. If $c_{\text{A}} \neq 0$, then $\dot h_{\text{A}}(\tilde u)$ approaches a nonzero constant at late times, and the integral on the R.H.S. acquires a term proportional to $c_{\text{A}}u$. Consequently, $h_{\text{A}}(u)$ would grow without bound as $u\to+\infty$, contradicting the assumed boundedness of the waveform. Hence,
\begin{equation}
\lim_{u\to+\infty}\dot h_{\text{A}}(u)=0.
\end{equation}
The same argument applied at early times gives
$\lim_{u\to-\infty}\dot h_{\text{A}}(u)=0$, and therefore
\begin{equation}
\Delta\dot h_{\text{A}}
\equiv
\dot h_{\text{A}}(+\infty)-\dot h_{\text{A}}(-\infty)
=0.
\end{equation}
Importantly, $\Delta\dot h_{\text{A}}=0$ does not imply $\Delta h_{\text{A}}=0$. A bounded waveform can approach different finite constants at early and late times, while its derivative approaches zero at both ends. For example,
$h_+(u)=A\tanh(\omega u)$ has $\Delta h_+=2A$ and $\Delta\dot h_+=0$. The conventional geodesic displacement memory can remain nonzero even though the spin-dependent contribution vanishes.

\noindent Hence, for bounded GW profiles for which $\dot h_{+,\times}$ have finite limits as $u\to\pm\infty$, the spin-dependent terms proportional to $\Delta\dot h_{+,\times}$ in Eqs.~(\ref{y_component_displacement_mem}) and~(\ref{z_component_displacement_mem}) do not contribute to the permanent displacement memory, whereas the conventional geodesic contribution proportional to $\Delta h_{+,\times}$ may remain nonzero. In other words, within the linear order approximation, spin-curvature coupling does not provide an additional contribution to the asymptotic
displacement-memory observable through the terms proportional to $\Delta\dot h_A$, although it can modify the transient response during the passage of the burst.

\noindent Moreover, as mentioned above, the vanishing of the spin-dependent contribution relies on the boundedness of the waveform. If this condition is relaxed, one can construct a waveform for which both the geodesic and spin-dependent contributions are nonzero. For example,
\begin{equation}
h_+(u)=A\tanh(\omega u)
+B\ln(\cosh(\omega u)),
\qquad
h_\times(u)=0.
\end{equation}
Substitution into Eqs.~(\ref{y_comp_devi_change}) and~(\ref{z_comp_devi_change}) then gives
\begin{align}
\Delta Y^y
&=-A Y^2_0+\frac{\omega Y^u}{m}B S_0^3, \qquad
\Delta Y^z =A Y^3_0-\frac{\omega Y^u}{m}B S_0^2,
\label{z_comp_displacement_displacement_mem}
\end{align}
where the first terms arise from the geodesic contribution and the second from the spin-curvature coupling. However, this profile is unbounded because $\ln(\cosh(\omega u))\sim |\omega u|-\ln 2$ as $|u|\rightarrow\infty$. Although both contributions can be made nonzero formally, this example lies outside the bounded-waveform class and, therefore, may not be an appropriate physical model for permanent GW displacement memory.

\section{Conclusion}\label{secV}
\noindent In this work, we have investigated the dynamics of spinning test particles in a weak plane GW spacetime, with emphasis on how intrinsic spin modifies their relative motion. We implement the MPD equations in the pole–dipole approximation, with the Tulczyjew spin supplementary condition, and use the corresponding worldline deviation equation to incorporate spin–curvature coupling into the detector response. These equations separate into the standard geodesic and additional terms generated by the MPD force that provides a systematic framework for identifying spin-dependent response, in particular, its implications for the gravitational memory effect. The main conclusions are as follows: 
\begin{itemize}
    \item For a plane GW propagating on a flat background (Minkowski), explicit calculation of the linearised curvature tensor allows us to determine the spin-induced corrections to the transverse deviation equations. Solving these equations perturbatively, we obtain the modified GW detector response, including the contribution from spin–curvature coupling. The spin-dependent correction introduces a mixing between the two GW polarizations and is proportional to the intrinsic spin and the longitudinal separation of the test particles, while the conventional geodesic response is recovered in the spinless limit.

    \item The tilted-detector analysis provides a consistent geometrical realisation of the finite longitudinal separation $Y^u$ required by the spin-dependent response. A rigid rotation of the interferometer preserves the equal arm lengths and orthogonality of the detector arms, further implying the detector response through the effective wave-to-detector transformation $M_{\text{eff}}=MR^{T}$. The tilt therefore provides a geometrical handle through which the spin–curvature contribution can, in principle, be probed.
    \item We next compute the displacement and velocity memory observables, exhibiting a spin-dependent contribution, with a finite nonzero $Y^u$, to the GW memory. In contrast to conventional displacement memory, which is governed by the asymptotic change in the GW strain, the spin-dependent contribution is controlled by asymptotic changes in the time derivatives of the waveform through curvature gradients entering the MPD force. Thus, the spin-dependent response is sensitive to the temporal structure of the GW burst in addition to its net strain change.

    \item Most importantly, for physically bounded waveforms, the spin-dependent contribution to the displacement memory vanishes at the linear order. For such waveforms $\dot h_{+,\times}\to 0$ must approach zero at both early and late times. Hence, $\Delta\dot h_{+,\times}=0$, and the spin–curvature coupling does not produce an additional asymptotic displacement-memory contribution, i.e., spin-induced memory effect, for bounded GW signals. However, it can modify the transient detector response during the passage of the wave.


\end{itemize}

\noindent Thus, the present analysis provides a first step toward a systematic study of the response of spinning probes to gravitational radiation and its possible implications for memory effects. Several natural extensions remain to be explored. It would be particularly interesting to investigate spin-dependent memory in exact plane-wave geometries, in a curved background. The formalism may also be generalized to include higher multipole moments, finite-size effects, and alternative spin supplementary conditions. In addition, it would be interesting to investigate whether the spin-dependent memory effects identified here admit a deeper connection with the asymptotic symmetries and the associated conserved charges \cite{Strominger:2017zoo}.

\noindent Overall, our study shows that spin–curvature coupling introduces a distinct contribution to the response of spinning probes to GWs, with a dependence on the temporal structure of the GW signal that is absent in conventional geodesic motion. This contribution does not provide a permanent displacement memory effect for physically bounded waveforms at linear order; however, it modifies the transient response during the passage of the wave and that can, in principle, provide a complementary probe of the radiative curvature structure. Thus, spinning particles may offer novel and intricate features of GW bursts/signal that are not directly encoded in conventional geodesic measurements. This can further enrich our understanding of the interplay between gravitational radiation, spacetime geometry and the dynamics of extended bodies in GR and beyond.

\section*{ACKNOWLEDGEMENTS}
\noindent R.A. and Shailesh Kumar (S.K.) acknowledges the support and research facilities provided by the Department of Physics, IIT Kharagpur.

\appendix

\section{Generalized antenna pattern functions for a tilted detector} \label{app:A}

\noindent The conventional (untilted) Michelson interferometer consists of two arms aligned along
$n_{(1)}$ and $n_{(2)}$, satisfying
$n_{(i)}\cdot n_{(j)}=\delta_{ij}$. The corresponding detector tensor is given by
\begin{align}
D^{\alpha\beta} =\frac{1}{2} (n^{\alpha}_{(1)}n^{\beta}_{(1)}-n^{\alpha}_{(2)}n^{\beta}_{(2)})
\label{untilted detector tensor}
\end{align}
Now, to construct the detector as a rigidly rotated plane out of the $yz$-plane, we demand that the arm vectors remain orthogonal, while the longitudinal coordinates of the two end masses are, in general, different. The beam splitter, located at the origin, remains fixed throughout the rotation. The detector consists of two orthogonal arms, each of length \(L\), and the entire detector plane is rotated rigidly by an angle \(\alpha\). Note that the detector plane cannot simply be rotated about one of its own arms, since such a rotation would treat the two interferometer arms asymmetrically. Instead, we choose the rotation axis to be
\begin{equation}
\hat{\mathbf a}
=
\frac{1}{\sqrt2}
(\hat y+\hat z).
\label{rot_axis}
\end{equation}
This rotation axis lies in the detector plane and bisects the angle between the two interferometer arms. Consequently, the detector plane is rotated rigidly through an angle $\alpha$ about the unit vector $\hat{\mathbf a}$, ensuring that both interferometer arms are treated symmetrically. Since the detector undergoes a rigid-body rotation, every vector associated with the detector transforms in the same manner. Accordingly, the unit vectors defining the tilted detector frame are given by
\begin{align}
n'_{(i)} = \mathbb{R}(\alpha)n_{(i)},\qquad i\in(1,2),
\label{tilted unit vectors}
\end{align}
or, equivalently, in component notation (index),
$n^{'\alpha}_{(i)} = R^{\alpha}{}_{\beta}n^{\beta}_{(i)}$. Here, $\mathbb{R}$ denotes the matrix representation of the rotation operator, while $R^{\alpha}{}_{\beta}$ denotes its components in index notation. The rotation matrix maps vectors from the original detector frame to the tilted detector frame. 

\noindent The rotation matrix $\mathbb{R}$ (or equivalently its components $R^{\alpha}{}_{\beta}$) is obtained using Rodrigues' rotation formula \cite{Goldstein2002, Murray1994},
\begin{equation}
\mathbb{R}(\alpha)
=
\mathbb{I}+(\sin\alpha)\mathbb{K}+(1-\cos\alpha)\mathbb{K}^2,
\end{equation}
where $\mathbb{K}$ denotes the skew-symmetric matrix associated with the rotation axis $\hat{\mathbf a}$,
\[
\mathbb{K}\equiv [\mathbf a]_\times
=
\begin{pmatrix}
0&-\frac1{\sqrt2}&\frac1{\sqrt2}\\
\frac1{\sqrt2}&0&0\\
-\frac1{\sqrt2}&0&0
\end{pmatrix},
\]
and $\mathbb{I}$ is the identity matrix. For the rotation axis defined above, the rotation matrix takes the explicit form
\begin{equation}
\mathbb{R}(\alpha)=
\begin{pmatrix}
\cos\alpha &
-\dfrac{\sin\alpha}{\sqrt{2}} &
\dfrac{\sin\alpha}{\sqrt{2}}
\\[2ex]
\dfrac{\sin\alpha}{\sqrt{2}} &
\dfrac{1+\cos\alpha}{2} &
\dfrac{1-\cos\alpha}{2}
\\[2ex]
-\dfrac{\sin\alpha}{\sqrt{2}} &
\dfrac{1-\cos\alpha}{2} &
\dfrac{1+\cos\alpha}{2}
\end{pmatrix}.
\label{eq:RodriguesMatrix}
\end{equation}
The rotation matrix $\mathbb{R}(\alpha)$ satisfies
$\mathbb{R}^{-1}=\mathbb{R}^{T}$,
$\mathbb{R}^{T}\mathbb{R}=\mathbb{R}\mathbb{R}^{T}=\mathbb{I}$,
and $\det(\mathbb{R})=1$. Using Eq.~(\ref{tilted unit vectors}), the unit vectors in the tilted frame are obtained as
\begin{align}
n'_{(1)} = \Big(-\frac{1}{\sqrt{2}}\sin\alpha,\frac{1}{2}(1+\cos\alpha),\frac{1}{2}(1-\cos\alpha)\Big), \hspace{0.2cm}
n'_{(2)} = \Big(\frac{1}{\sqrt{2}}\sin\alpha,\frac{1}{2}(1-\cos\alpha),\frac{1}{2}(1+\cos\alpha)\Big).
\end{align}
The rotated unit vectors continue to satisfy the orthonormality condition,
$n'_{(i)}\cdot n'_{(j)}=\delta_{ij}$,
as in the untilted frame. 

\noindent Using these rotated unit vectors, we define the detector tensor in the tilted frame as
\begin{align}
D'^{\alpha\beta} =\frac{1}{2}(n'^{\alpha}_{(1)}n'^{\beta}_{(1)}-n'^{\alpha}_{(2)}n'^{\beta}_{(2)}).
\end{align}
Substituting Eq.~(\ref{tilted unit vectors}), the tilted detector tensor can be written as
\begin{align}
D'^{\alpha\beta} = R^{\alpha}{}_{\gamma}R^{\beta}{}_{\delta}D^{\gamma\delta},
\end{align}
where $D^{\alpha\beta}$ is the detector tensor in the original (untilted) frame, given by Eq.~(\ref{untilted detector tensor}). Equivalently, in matrix notation, this relation can be written as
$D'=\mathbb{R}D\mathbb{R}^{\text{T}}$.
This expression shows that the rotation modifies only the orientation of the detector, while its intrinsic geometry remains unchanged. Now, using this detector tensor, let us next construct the generalised antenna pattern functions for the tilted detector.

\subsection{Relation between the wave frame and the tilted detector frame}

\noindent We now establish the relation between the tilted detector frame and the wave frame, which is required to determine the generalised antenna pattern functions for the rotated detector configuration. Let us introduce the wave frame
$W=\{\hat{e}_{X},\hat{e}_{Y},\hat{e}_{Z}\}$, where $\hat{e}_{X}$ denotes the direction of wave propagation, and $\{\hat{e}_{Y},\hat{e}_{Z}\}$ span the transverse polarisation plane. The corresponding polarisation tensors are given by
\begin{align}
e^{+}_{ij} = \hat{e}_{Y_{i}} \hat{e}_{Y_{j}} - \hat{e}_{Z_{i}} \hat{e}_{Z_{j}}, \qquad e^{\times}_{ij} = \hat{e}_{Y_{i}} \hat{e}_{Z_{j}} - \hat{e}_{Z_{i}} \hat{e}_{Y_{j}},   
\end{align}
where ($i,j$) denote spatial Cartesian indices. The original detector frame is denoted by $D=\{\hat{e}_{x},\hat{e}_{y},\hat{e}_{z}\}$. Here, we switch from the notation $n_{(i)}$ used for the detector arms to $\hat{e}_{i}$ for the detector basis vectors (for notational convenience), with the corresponding primed quantities representing the tilted detector frame. The wave frame and the original detector frame are related through the Euler transformation matrix \cite{Dhurandhar:2022eik}, which for the conventional detector is written as
\begin{align}
    \hat{e}_{A} = M_{A}{}^{\alpha}\hat{e}_{\alpha},
\end{align}
where the index $A$ labels the wave frame basis vectors ($X, Y, Z$), and $\alpha$ denotes the detector frame basis vectors ($x, y, z$). In the tilted detector frame, the basis vectors are related through the rotation matrix as
\begin{align}
    \hat{e}_{\alpha'} = R_{\alpha'}{}^{\beta}\hat{e}_{\beta}, \qquad \hat{e}_{\alpha} = (R^{\text{T}})_{\alpha}{}^{\beta'}\hat{e}_{\beta},
\end{align}
where $R^{\text{T}}$ denotes the transpose of the rotation matrix. Combining the above relations, we obtain
\begin{align}
    \hat{e}_A = M_{A}{}^{\alpha} (R^{\text{T}})_{\alpha}{}^{\beta'} \hat{e}_{\beta'} =  (M R^{\text{T}})_{A}{}^{\beta'} \hat{e}_{\beta'}.
\end{align}
Thus, the wave frame is directly related to the tilted detector frame through the effective transformation
$M_{\mathrm{eff}}=MR^{\mathrm{T}}$. The elements of this effective transformation matrix are:
\begin{align}
    M_{\text{eff}}\equiv m'_{A_{i}} = M_{A_{j}} (R^{\text{T}})_{ji} = 
\begin{pmatrix}
m'_{11} & m'_{12} & m'_{13} \\
m'_{21} & m'_{22} & m'_{23} \\
m'_{31} & m'_{32} & m'_{33}
\end{pmatrix},
\end{align}
where
\begin{align}
    m'_{11} =& \cos\alpha M_{11} + \frac{\sin\alpha}{\sqrt{2}} (M_{13}-M_{12}) \\
    m'_{12} =& \frac{\sin\alpha}{\sqrt{2}}M_{11}+\frac{1}{2}(1+\cos\alpha)M_{12}+\frac{1}{2}(1-\cos\alpha)M_{13} \\
    m'_{13} =& -\frac{\sin\alpha}{\sqrt{2}}M_{11}+\frac{1}{2}(1-\cos\alpha)M_{12}+\frac{1}{2}(1+\cos\alpha)M_{13} \\
    m'_{21} =& \cos\alpha M_{21} - \frac{\sin\alpha}{\sqrt{2}} (M_{22} -  M_{23}) \\
    m'_{22} =& \frac{\sin\alpha}{\sqrt{2}}M_{21} +\frac{1}{2}(1+\cos\alpha)M_{22}+\frac{1}{2}(1-\cos\alpha)M_{23} \\
    m'_{23} =& -\frac{\sin\alpha}{\sqrt{2}}M_{21} +\frac{1}{2}(1-\cos\alpha)M_{22}+\frac{1}{2}(1+\cos\alpha)M_{23} \\
     m'_{31} =& \cos\alpha M_{31} - \frac{\sin\alpha}{\sqrt{2}} (M_{32} -  M_{33}) \\
    m'_{32} =& \frac{\sin\alpha}{\sqrt{2}}M_{31} +\frac{1}{2}(1+\cos\alpha)M_{32}+\frac{1}{2}(1-\cos\alpha)M_{33} \\
    m'_{33} =& -\frac{\sin\alpha}{\sqrt{2}}M_{31} +\frac{1}{2}(1-\cos\alpha)M_{32}+\frac{1}{2}(1+\cos\alpha)M_{33},
\end{align}
where the coefficients $M_{ij}$ are of the Euler matrix for the wave propagating in $x$-direction, which can be easily found in\cite{Goldstein2002, Dhurandhar:2022eik}.

\noindent With the effective transformations in hand, to determine the antenna pattern functions for the tilted detector, we first evaluate the following scalar product:
\begin{align}
    \hat{e}_{A}\cdot \hat{e}_{\alpha'} = [(M R^{\text{T}})_{A}{}^{\beta'}\hat{e}_{\beta'}] \cdot \hat{e}_{\alpha'} = (M_{\text{eff}})_{A\alpha'}
\end{align}
$\hat{e}_{Y}\cdot \hat{e}_{y'}=m'_{22}$, $\hat{e}_{Y}\cdot \hat{e}_{z'}=m'_{23}$, $\hat{e}_{Z}\cdot \hat{e}_{y'}=m'_{32}$ and $\hat{e}_{Z}\cdot \hat{e}_{z'}=m'_{33}$.
Using these relations, the generalised antenna pattern function for the plus polarization is obtained as
\begin{align}
    F'_{+} \equiv e^{+}_{ij} D'_{ij} =& \frac{1}{2}(\hat{e}_{Y_{i}} \hat{e}_{Y_{j}} - \hat{e}_{Z_{i}} \hat{e}_{Z_{j}}) (\hat{e}^{i}_{y'} \hat{e}^{j}_{y'}-\hat{e}^{i}_{z'} \hat{e}^{j}_{z'}) \\
    =& \frac{1}{2}[(\hat{e}_{Y}\cdot \hat{e}_{y'})^2 - (\hat{e}_{Y}\cdot \hat{e}_{z'})^2 - (\hat{e}_{Z}\cdot \hat{e}_{y'})^2 + (\hat{e}_{Z}\cdot \hat{e}_{z'})^2] \\
    =& \frac{1}{2}[m'^{2}_{22} - m'^{2}_{23} - m'^{2}_{32} + m'^{2}_{33}]
\end{align}
Similarly, the generalised antenna pattern function for the cross polarization is given by
\begin{align}
    F'_{\times} \equiv e^{\times}_{ij} D'_{ij} =&  \frac{1}{2}(\hat{e}_{Y_{i}} \hat{e}_{Z_{j}} + \hat{e}_{Z_{i}} \hat{e}_{Y_{j}}) (\hat{e}^{i}_{y'} \hat{e}^{j}_{y'}-\hat{e}^{i}_{z'} \hat{e}^{j}_{z'}) \\
    =& (\hat{e}_{Y}\cdot \hat{e}_{y'}) (\hat{e}_{Z}\cdot \hat{e}_{y'}) - (\hat{e}_{Y}\cdot \hat{e}_{z'}) (\hat{e}_{Z}\cdot \hat{e}_{z'}) \\
    =& m'_{22} m'_{32}-m'_{23} m'_{33}.
\end{align}
Therefore, the tilted detector geometry modifies the antenna response by introducing an explicit dependence on the detector tilt angle $\alpha$, in addition to the usual dependence on the Euler angles $(\Phi,\Theta,\Psi)$ that specify the relative orientation between the source and the detector. Accordingly, generalised antenna pattern functions can be expressed as
\begin{equation}
F'_{+,\times} \equiv F'_{+,\times}(\Phi,\Theta,\Psi;\alpha).
\end{equation}
Consequently, the measured detector strain inherits the same functional dependence on the Euler angles and the detector tilt angle. 

\noindent As a consistency check, in the limit $\alpha\rightarrow0$, the rotation matrix becomes the identity matrix, i.e., $\mathbb{R}=\mathbb{I}$, and hence, $M_{\rm eff}=\mathbb{M}$. Therefore,
$F_{+} \rightarrow \frac{1}{2}(M_{22}^{2} - M_{32}^{2} - M_{23}^{2} + M_{33}^{2})$, and $F_{\times} \rightarrow \left(
M_{22}M_{32} - M_{23}M_{33} \right)$, implying that the effective transformation matrix reduce to the Euler transformation matrix, and the generalized antenna pattern functions reduce to the standard
ones. 

\bibliographystyle{apsrev}
\bibliography{bibfile}

\appendix

\end{document}